# Evaporation of bacteria-laden surrogate respiratory fluid droplets: On a hydrophilic substrate versus contact-free environment confers differential bacterial infectivity

Amey Nitin Agharkar[1#] Dipasree Hajra[2#], Durbar Roy[3#], Vivek Jaiswal[3], Prasenjit Kabi[3], Dipshikha Chakravortty[2,4] and Saptarshi Basu[1,3*]

[1] Interdisciplinary Centre for Energy Research (ICER), Indian Institute of Science

[2] Department of Microbiology & Cell Biology, Indian Institute of Science

[3] Department of Mechanical Engineering, Indian Institute of Science.

[4]Adjunct Faculty, School of Biology, Indian Institute of Science Education and Research, Thiruvananthapuram

[#] Equal First Authors

Authorship Note – ANA and DH have contributed equally to this work.

Conflict of Interest – The authors have declared that no conflict of interest exists.

*Corresponding authors: Prof. Saptarshi Basu

Email: sbasu@iisc.ac.in

# Abstract

The transmission of viruses/ bacteria cause infection predominantly via aerosols. The transmission mechanism of respiratory diseases is complex, including direct or indirect contact, large droplet, and airborne routes apart from close contact transmission. With this pretext, we have investigated two modes of droplet evaporation to understand its significance in airborne disease transmission; a droplet in a contact-free environment, which evaporates and forms droplet nuclei, and a droplet on a hydrophilic substrate (fomite). The study examines mass transport, the deposition pattern of bacteria in the precipitates, and their survival and virulence. The osmotic pressure increases with the salt concentration, inactivating the bacteria embedded in the precipitates with accelerated evaporation. Further, the bacteria's degree of survival and enhanced pathogenicity are compared for both evaporation modes. The striking differences in pathogenicity are attributed to the evaporation rate, oxygen availability, and reactive oxygen species (ROS) generation.

# Introduction

Human respiratory diseases encompass a vast range of disease severity and symptoms, leading to widespread mortality, morbidity, and economic loss worldwide. Most respiratory diseases are caused by various viruses like influenza, SARS, MERS, or coronaviruses. [1] However, several viral infections predispose the host toward secondary bacterial infections. [2] Bacteria like *Klebsiella pneumoniae* and *Pseudomonas aeruginosa* enter the host through the nasal passage and cause pneumonia, and they are the leading causes of nosocomial respiratory infections in humans. [3,4] There exists a positive correlation between viral respiratory tract infections and bacterial infections in rhinitis, RSV-induced bronchiolitis, acute respiratory wheezing, and cystic fibrosis, wherein *Pseudomonas aeruginosa* and *Staphylococcus aureus* are the major causative agents of the respiratory exacerbations. [1] Such bacterial infections trigger severe chronic obstructive pulmonary disease (COPD) exacerbations. A recent case report highlighted non-typhoidal *Salmonella* (NTS) bacteraemia in a 34-year-old immunocompetent Japanese male patient with moderate COVID-19 post methylprednisolone discontinuation. [5] *Mycobacterium tuberculosis* and non-tuberculous bacteria are leading causes of lung diseases worldwide. The incidences of non-tuberculous mycobacterial lung diseases have increased over the past decades in several parts of the world, including the United States. [6,7]

Humans coughing, sneezing, and speaking produce a jet of aerosols consisting of many microdroplets with moist air. [8] Droplets emitted by an infected human containing bacteria/ viruses are considered as primary routes of transmitting respiratory disease [9] to susceptible individuals via four major modes, namely direct contact, indirect contact (fomites), large droplets, or fine aerosols. [10-16] Emitted droplets based on their initial diameters were dichotomized by Wells. [17,18] Smaller droplets (< 100 μm) will travel a certain distance while evaporating in air, and the bigger ones will settle down quickly within a few seconds of emission. It is assumed by many investigators that the evaporation of small droplets occurs very quickly, so the evaporation kinetics and dynamics of such droplets are unknown. [19] The evaporation process of the emitted droplet in the environment can be simulated by a single droplet experiment in an acoustic levitator because the relative velocity-based Reynolds number between droplets and surrounding droplet jets reduces to a small value very quickly. [20] It is globally observed and confirmed, for a range of droplet diameters, that droplet nuclei (precipitate) size after droplet evaporation is 20-30% of the initial size and is independent of initial droplet size. [20] After evaporation of the emitted droplets, the nuclei can enter the host's respiratory tract through inhalation and can cause lung diseases. Droplets settling down on the fomite surface are also significant modes of transmission of microbial infection. [21]

Finally, the size of the inhaled/ingested droplet plays a significant role in pathogenesis. For example, a particle size below 5μm is believed to settle in the pulmonary alveolar region of the lung and cause infection, whereas a particle larger than 5μm can get deposited onto the nasal passage via the centrifugal force. [22,23] Other studies have suggested that a particle smaller than 10μm can infect the pulmonary region,[24] and a particle of size range between 10μm and 100μm can get inhaled and deposited in the head airways or the tracheobronchial regions. [14]

Studying the aerosol jet in the bulk phase is difficult because it has a size range of microdroplets with moist air and conditions such as supersaturation and high surface area to volume ratio are unique. [25] Since a droplet is an integral part of aerosol, analysing a single droplet can answer many questions about the physics of evaporation and pathogenicity of droplet-carrying bacteria. The levitator offers a controlled methodology for studying the host of dynamics, such as mass transfer, crystallization, or droplet breakup. [26]

Our study demonstrates the role of the differential precipitation method in conferring virulence to infectious pathogens. We have compared the levitated precipitates and the sessile precipitates maintaining the same initial volume of bacterial solution, aiming to deduce the effect of evaporation on bacterial survival and virulence. Therefore, our study was performed by producing bacteria-laden surrogate respiratory fluid (see methods) drops of diameter 650 ± 20 μm, and the ambient condition was maintained by complying with the tropical climate condition (29 °C ± 2 °C, 42 ± 3 % RH). An increase in the salt concentration (solute) after the evaporation of droplets tends to inactivate the bacteria in precipitates due to an increase in osmotic pressure.[19] Here we show that the levitated precipitate harbour more viable and infectious bacteria than its sessile precipitate counterpart owing to increased ROS. In this study, we revealed which precipitate, levitated or sessile, is more virulent, and the reason for its virulence is attributed to the evaporation rate and ROS.

# Experimental Methods

## Surrogate respiratory fluid preparation

The constituents of the surrogate respiratory fluid (SRF) were 0.9 % by wt. of NaCl, 0.3 % by wt. of gastric mucin (Type III, Sigma Aldrich), and 0.05 % wt. of di-palmitoyl-phosphatidyl-choline (DPPC (Avanti Polar Lipids)) in deionized water. [27] The final formulation was sonicated (TRANS-O-SONIC sonicator) for 15 minutes and centrifuged (REMI R-8C centrifuge) at 5000 rpm for 15 minutes. [28]

# Experimental setup for levitated and sessile droplet evaporation

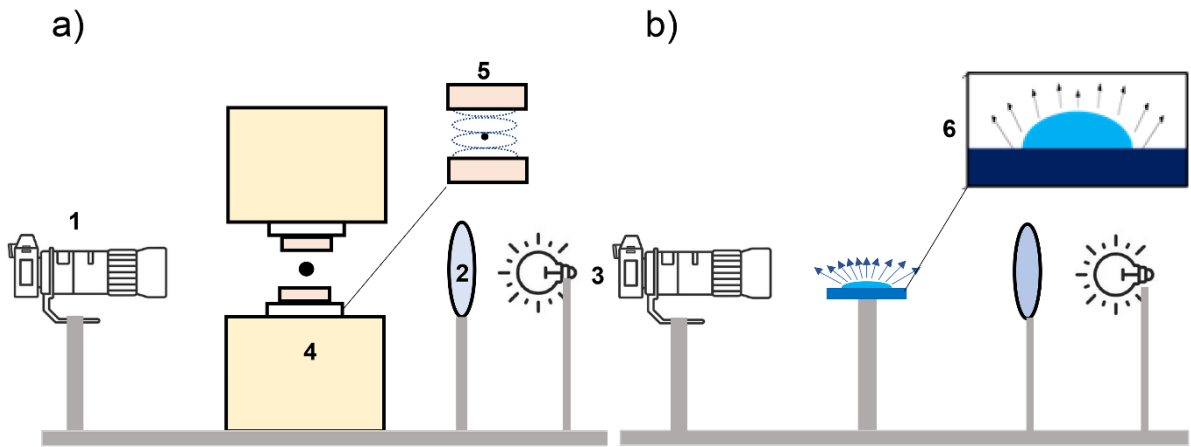

**Figure 1: Experimental setup for a) Levitated droplet 1) camera 2) diffuser plate 3) light source 4) acoustic levitator 5) droplet in levitated condition; b) Sessile droplet 6) sessile droplet on a glass slide.**

All experiments were performed at an ambient temperature of 29 °C ± 2 °C and RH of 42 ± 3 % as relative humidity (RH) (measured using TSP-01 sensor, Thorlabs). Figure 1 a) shows the experimental setup. The droplet is levitated using an ultrasonic levitator (tec5). A commercial DSLR camera (Nikon D5600) fitted with a Navitar zoom lens assembly (2 X lens × 4.5 X tube) is used to capture the lifetime of the levitated droplet at 30 frames/second and a spatial resolution of 1 pixel/μm. A diffuser plate is inserted between a LED light source (5W, Holmarc) and the acoustic levitator for uniform illumination. The droplet of initial diameter 650 ± 20 μm is created using a DISPO VAN insulin syringe (31G, 0.25 × 6 mm) of 1 mL and placed on the levitator node as shown in Figure 1 a). The levitated droplet has an elliptical appearance due to higher pressure at its polar regions. The effective diameter was estimated as $D = \sqrt[3]{d_x^2 d_y}$, where $d_x$ is the major diameter of the droplet and $d_y$ is the minor diameter of the droplet. The temporal evolution of the droplet diameter during evaporation was extracted using the "Analyse particles" plugin of Image J (open-source image processing software). Post-evaporation, the levitated precipitate was carefully collected on microscopic cover glass (BLUE STAR, square 22 mm).

Figure 1 b) 6) represents the evaporation of the sessile droplet. At the same time, the setup used was the same. Glass slides (BLUE STAR micro slides 75mm × 25 mm) were sonicated for 2-4 minutes using propan-2-ol bath and were wiped by Kimwipes (Kimberley Clark International). DISPO VAN insulin syringe (31G, 0.25 × 6 mm) of 1 mL generated a droplet of 0.15 ± 0.02 μL volume equivalent to that of the levitated droplet and was gently placed on the glass slide. The side view of the drying droplet was imaged using the same setup as used for the levitated droplet. Simultaneously, the top view was recorded using an Olympus optical microscope at a 0.4 pixel/μm spatial resolution. As the Capillary number ($Ca = \mu u/\gamma \sim 10^{-8}$) and Bond number ($Bo = (\rho g R_c h/\gamma) \sim 10^{-2}$) comply with the spherical cap model, the volume of the sessile droplet was estimated using $V_s = (\pi h/6)(3R_c^2 + h^2)$ where $R_c$ is the contact radius, and h is the drop height at center. [29] The temporal evolutions of the height and radius of the evaporating sessile droplet were extracted from Image J's "Analyse particles" plugin.

**Bacterial sample preparation:** Overnight grown stationary phase cultures of *Salmonella* Typhimurium (STM) 14028S, *Pseudomonas aeruginosa* PA01, *Mycobacterium smegmatis* mc$^2$ 155, and *Klebsiella pneumoniae* MH1698 were taken, and their absorbance was measured at OD600nm. $10^9$ CFU (Colony Forming Units) of the bacterial culture were pelleted down at 6000 rpm for 6 minutes. The pellets were washed twice with phosphate-buffered saline (PBS) pH 7.0 and finally resuspended in 500 µL Surrogate respiratory fluid (SRF). For confocal studies, *S.* Typhimurium expressing mCherry (pFV-mCherry (RFP)) (STM-RFP) and *S.* Typhimurium expressing Green Fluorescent Protein (GFP) (pFV-GFP) (STM-GFP) strains were used.

## Confocal microscopy and analysis

*S.* Typhimurium expressing mCherry (pFV-mCherry (RFP)) (STM-RFP), *Pseudomonas aeruginosa* PA01, *Mycobacterium smegmatis* mc$^2$ 155, and *Klebsiella pneumoniae* MH1698 samples were prepared in 500µL SRF and were subjected to the levitated and sessile mode of evaporation as mentioned before. All other bacterial strains except STM-RFP were stained with FM4-64 dye (1µg/mL) for 15-20 min at 37ºC protected from light before resuspension in 500 µL SRF fluid. Zeiss LSM 880 NLO upright multi-photon confocal microscope was used to image the levitated and sessile samples precipitates at 10X magnification. ZEN Black software (Carl Zeiss) was used to obtain the Z stacks images to study the bacterial deposition in the precipitates.

**In vitro bacterial viability assessment:** The levitated and the sessile precipitated droplet were reconstituted in 10µL of sterile, autoclaved PBS. 990µL of sterile PBS was added to the 10μL retrieved droplet precipitate, and 100μL of it was plated on LB agar plates. For *M. smegmatis* growth LB containing 0.01% Tween 80 agar plates were used. Similarly, 100μL of the bacterial sample ($10^9$ CFU/mL in 500μL SRF) were plated onto the respective plates at dilution of $10^{-7}$ and $10^{-8,}$ which served as the pre-inoculum for normalization by considering the volume of the droplet (approximately 0.15μL). 16h post plating and incubation at 37ºC incubator, the bacterial count was enumerated. The survival percentage was calculated as follows:

$$\frac{\text{Number of viable colonies on the sessile or levitated precipitate}}{\text{Number of viable colonies in the pre-inoculum before evaporation}} \times 100$$

**Cell Culture and Infection studies:** RAW264.7 murine macrophages were used for the infection studies. The cells were cultured in DMEM (Lonza) containing 10% Fetal Bovine Serum (Gibco) at 37 ºC in a humified incubator with 5% $CO_2$. 12h-14h prior to each experiment; cells were seeded into 24 well at a confluency of 60%. The bacteria reconstituted from the levitated or sessile precipitated droplet were used for infecting RAW264.7 macrophages. The sessile and levitated reconstituted precipitates were mixed in a 1:1 ratio to infect the macrophages for the mixed culture experiment. The infected cells were incubated in 5% $CO_2$ incubator for 25min at 37ºC for *S.* Typhimurium (STM) and *P. aeruginosa* (PA) infection and for *K. pneumoniae* (KP) and *M. smegmatis* (MS) infection, the infected cells were incubated for 1h and 2h respectively. Post specific incubation, the infected cells were washed with sterile PBS and were incubated at 37ºC in a 5% $CO_2$ incubator for 1h in the presence of 100μg/mL gentamicin containing DMEM media. Post 1hr incubation, cells were further incubated with 25μg/mL gentamicin containing DMEM media at 37ºC in a 5% $CO_2$ incubator after a PBS washing step. At designated time points post-infection, (2h,16h-STM, PA, KP;4h,24h-MS) cells were lysed with 0.1% Triton-X-100 in PBS. The lysates were serially diluted and plated on LB agar plates. The CFU at the initial time-point of infection was divided by the respective pre-inoculum CFU to obtain the percent phagocytosis. Additionally, the CFU at the later time points of infection (y= 16h for STM, PA, and KP; y =24h for MS) was divided by the corresponding CFU at the initial time points of infection (x=2h for STM, PA, and KP; x=4h for MS) to calculate the fold proliferation.

$$\textbf{Percentage Phagocytosis} = \frac{[\textbf{CFU at x hours}]}{[\textbf{CFU of pre}-\textbf{inoculum}]}$$

$$\textbf{Fold proliferation} = \frac{[\textbf{CFU at y hours}]}{[\textbf{CFU at x hours}]}$$

**Confocal Microscopy**

For the confocal imaging experiment, cells were seeded onto coverslip laden 24 well plate at a confluency of 60% one day prior to the infection. The infection protocol was followed as described previously. The cells were washed in PBS and fixed in 3.5% paraformaldehyde at the specific time points post-infection with STM-RFP or STM-GFP reconstituted from their respective levitated or sessile precipitated droplet. The coverslips were mounted onto a clean glass slide using the mounting media containing the anti-fade agent. The coverslip sides were sealed with a transparent nail paint and were subsequently subjected to imaging using Zeiss LSM 880 multiphoton confocal microscope using a 63X objective. Image analysis was performed using ZEN Black 2012 software.

**Flow Cytometry:** The sessile and the levitated precipitate droplets were reconstituted in 1X PBS. The reconstituted precipitates were centrifuged at 6000rpm for 6 minutes to obtain the bacterial pellet. The pellets were resuspended in the DCFDA (2',7'Dichlorofluorescein diacetate) (20µM) staining solution and incubated for 15 mins in the dark at 37ºC. For Propidium Iodide (PI) staining, the cell pellets were resuspended in 1X PBS PI staining solution at a concentration of 1µg/ml, and the incubation was performed in the dark at room temperature. Following incubation, centrifugation was performed at 6000 rpm for 6 minutes to obtain the stained bacterial pellet. The bacterial pellets were washed in 1X PBS and eventually resuspended in 100µL of 1X PBS for FACS recording. The FACS recording was performed in BDFacsVerse instrument, and analysis was performed in BDFacsSuite software.

**Statistical analysis**: Data were analyzed, and graphs were plotted using the GraphPad Prism 8 software (San Diego, CA). Statistical significance was determined by Student's t-test to obtain p values. Adjusted p-values below 0.05 are considered statistically significant. The results are expressed as mean ± SEM.

**Bacterial fluid property measurement:** Averaged Viscosity was measured by Anton Paar TwinDrive Rheometer (MCR 702) at the shear rate range of 10-1000 1/s at 25 ºC. Density and Surface tension data was acquired at 29 ± 2 ºC temperature and 42 ± 3% RH. Density was measured using the ratio of mass and volume of the sample. Surface tension was measured using pendant droplet technique and image analysis was done by "ImageJ" software.

# Results and Discussion

## Evaporation dynamics of levitated and sessile droplet

We began with a comparative evaporation study of bacteria-laden droplets in a contact-free environment and on a hydrophilic substrate. The contact-free environment mode of droplet evaporation was mimicked by a droplet in the levitated condition in an acoustic levitator. In contrast, the droplet on a hydrophilic substrate was demonstrated by the droplet in a sessile mode of evaporation on a glass slide (see Methods). In this work, we show that a higher fraction of bacteria survives (hence are more infectious) in levitated configuration than in sessile configuration during drop evaporation for a fixed initial drop volume. This important result is independent of drop size for the same volume of drop used in both sessile and levitated configuration. This could be understood based on the fact that given an initial drop volume, the evaporation dynamics will be similar for both small and large drops provided the drop size is in the diffusion limited evaporation phase which is the case for the present work. The most important parameter in such drop evaporation scenarios is the percentage volume reduction/mass loss from the evaporating drop. The percentage drop volume reduction is independent of drop size and depends majorly on physical conditions/parameters and material properties like evaporation flux, drop temperature, ambient temperature, relative humidity, vapor concentration field (vapor pressure), flow field inside and outside the drop, thermal conductivity and diffusivity, mass diffusivity, solute properties to name a few. Further, the evaporation time is also directly proportional to the percentage volume reduction.

Inset I in Figure 2(a) show a slightly flattened droplet due to the acoustic pressure at the poles. [30,31] For the sake of brevity, cases are referred to by their composition. Thus, droplets comprising only surrogate respiratory fluids (see Methods) are called SRF. SRF droplets containing pathogens (see Methods) are referred to by the abbreviation of the pathogen, e.g., *M. smegmatis* (MS), and so on.

Inset (II) in Figure 2 still shows the drop as spherical. From Figure 2 we can observe that the evaporation rate slows down as time progresses for the levitated configuration in contrast to the constant evaporation rate for the sessile configuration. At the end of evaporation, we can observe precipitate formation in SRF at around t ~500-550 s. As the droplet approaches the efflorescence limit, it undergoes rapid shape changes. Efflorescence is the onset of crystallization at a particular supersaturation concentration. This results in significant errors while estimating the droplet diameter. The end of evaporation is confirmed by the transition of the droplet shape from near-spherical to irregular with jagged edges (see Inset III in Figure 2) and the formation of a fully evaporated precipitate. We can observe the evolution of volume ratio as a function of time in both levitated and sessile condition for STM-WT and SRF in Figure 2. It is important to note that the difference in evaporation time for a fixed drop volume is drastic for the levitated and sessile configuration. The difference in evaporation time between levitated and sessile mode of evaporation emanates majorly from the geometry as is demonstrated below.

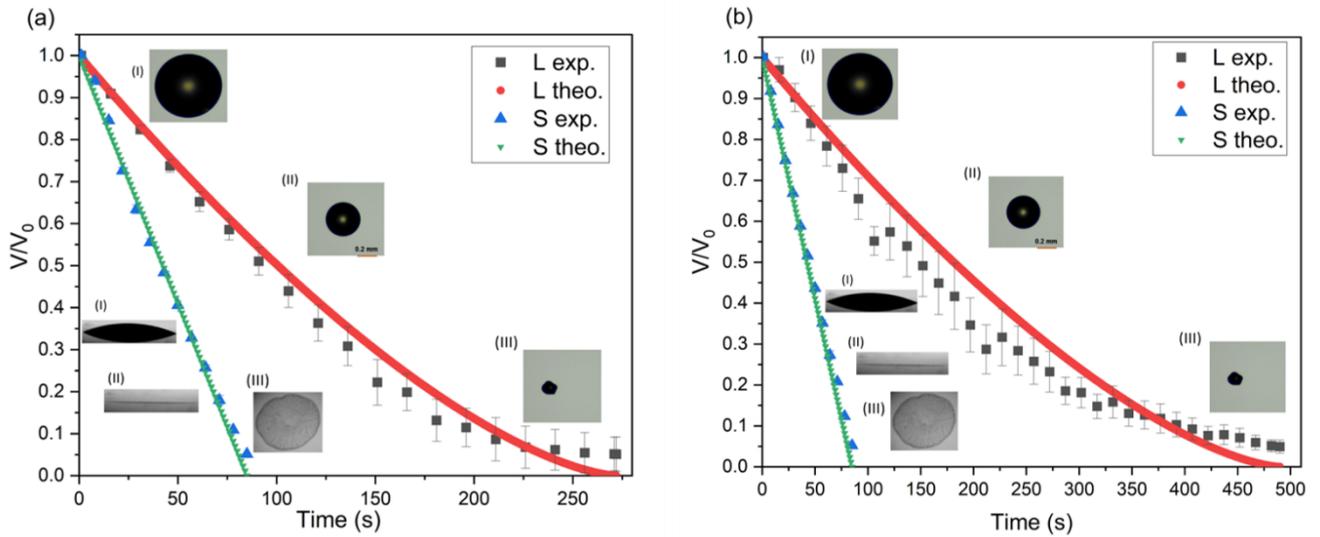

**Figure 2:** Volume ratio $V/V_0$ variation as a function of time in seconds for levitated and sessile drop evaporation for (a) STM WT and (b) SRF. The theoretical model is shown is solid line and the experimental data is shown using solid markers. Here L, S denotes levitated and sessile configuration respectively. The keywords exp. and the. represents the experimental and theoretical scales of the volume regression.

From Figure 2 we can observe that the evaporation time scale is of the order O (100 s) for sessile and O (250-500 s) for levitated configuration depending on drop size and relative humidity. The evaporation process being slow can be fully understood by diffusion driven quasi-steady evaporation model. The length scale that determines whether evaporation occurs within the diffusion or reaction limited regime is given by the ratio of water diffusivity to the water vapor condensation rate constant obtained in terms of the molecular velocity at a given temperature (i.e., $D/k_c$) where $D$ is the water vapor diffusivity and $k_c$ is the condensation rate constant. In general, the condensation rate constant is proportional to the molecular speed of a water molecule in the gas phase (i.e. $k_c \sim \sqrt{k_B T/m}$, where $k_B$ is the Boltzmann constant, $T$ is the ambient temperature of the surrounding air, $m$ is the mass of a water molecule). Owing to the fact that within our current experiments the smallest drop size that we attain during the end of evaporation is of the order of 10 microns which is at least two orders of magnitude larger than the length scale 0.1 microns where reaction limited evaporation dynamics occurs. Therefore, in the context of the current work, the entire evaporation process occurs in the diffusion limited regime.

Assuming spherical symmetry for a drop of radius $R_l$ (subscript '$l$' denotes acoustic levitator) the rate of mass loss from the drop surface due to evaporation is given by [32]

$$\frac{dm_l}{dt} = -2\pi \rho_a D R_l Sh^* ln(1 + B_M)$$

(1)

where $m_l = (4/3)\rho_f \pi R_l^3$ represents the instantaneous drop mass for a fluid density $\rho_f$. $\rho_a$ is the air density, $D$ is the water vapor diffusivity, $Sh^*$ represents the modified Sherwood number calculated as $Sh^* = 2 + (Sh_0 - 2)/F_M$. $B_M$ is the Spalding Mass transfer number is computed as $B_M = c_0(1 - RH)/(\rho_a - c_0)$ where $c_0$ is the saturation concentration of water vapor on the droplet surface and $RH$ is the relative humidity. $F_M$ being a function $B_M$ is calculated as $F_M = (1 + B_M)^{0.7} \ln(1 + B_M)/B_M$. The term $F_M$ used to calculate $Sh^*$ quantifies the effect of Stefan flow on determining a modified Sherwood number used to calculate the evaporative mass loss. $Sh_0 = 2 + 0.522 Re^{1/2} Sc^{1/3}$ where $Re = (\rho_a v R_{l0})/\mu_a$ is the Reynolds number where $v$ is a velocity scale in the surrounding air, and $Sc$ is the Schmidt number given by $Sc = \nu/D$ where $\nu$ is the kinematic viscosity. The effect of surrounding air flow on the evaporation rate becomes dominant when the flow velocity has non-zero component along the evaporative flux direction (radial direction in the present case) and the functional dependence is given by the relationship of Sherwood number on the Reynolds number as discussed above. For the present case of a drop evaporating in an acoustic levitator, evaporation primarily occurs due to water vapor concentration gradient in the radial direction. The mean flow velocity in the radial direction is negligible (as majority of the flow field is tangential close to the drop surface) owing to closed streamlines in the velocity field of an acoustic levitator. Therefore, we can use $Sh^* = Sh_0 = 2$ and the evaporation becomes decoupled from the surrounding flow field and is diffusive in nature. Substituting the value of $m_l$ and Sherwood number in the above equation, we have

$$R_l \frac{dR_l}{dt} = -\frac{F}{\rho_f} \tag{2}$$

where $F = \rho_a D \ln(1 + B_M)$.

Integrating the above equation recovers the familiar $R^2$ law

$$R_l^2 = R_{l0}^2 - \frac{2Ft}{\rho_f} \tag{3}$$

Taking the square root on both sides of equation (3), the drop radius as a function of time becomes

$$R_l = \sqrt{R_{l0}^2 - \frac{2Ft}{\rho_f}} \qquad (4)$$

The instantaneous drop volume represented as a fraction of initial volume ($V_l(t)/V_{l0}$) therefore becomes

$$\frac{V_l(t)}{V_{l0}} = \left(1 - \frac{t}{t_{eva,l}}\right)^{3/2} \qquad (5)$$

as volume is proportional to the cube of drop radius, where

$$t_{eva,l} = \frac{\rho_f R_{l0}^2}{2\rho_a D \ln(1 + B_M)} \qquad (6)$$

The drying of non-volatile solute laden water drops proceeds in two distinct stages. Stage one consists of solvent (water) evaporation from drop surface which forms major percentage of drop evaporation time and stage two occurs at the end stage of evaporation when solute concentration increases. During stage two, the evaporation rate considerably reduces as a result of suppression of vapor pressure due to increase in solute concentration according to Raoult's law at the drop surface. The slowing down of evaporation rate coupled with the increase in non-volatile solute concentration results in rapid precipitate formation and drop size remains constant.[33] The actual evaporation time scale can therefore deviate from the above expression due to other higher order effects like non-volatile solute induced water vapor pressure reduction, evaporative cooling, concentration gradient inside the drop and precipitate formation as was shown by Rezaei and Netz. [34,35] However, it is important to note that for small solute concentration as for the present case, the drop volume regression in an acoustic levitator is accurately given by the scale $(1 - t/t_{eva,l})^{3/2}$ for the major part of drop regression process. $t_{eva,l}$ is experimentally measured as the time a drop takes to attain a volume ratio of 0.05. Figure 2(a) and 2(b) shows a comparison of the drop evaporation theoretical model (red solid curve) with the experimental data for STM WT and SRF. The experimental volume regression ratio agrees with the theoretical scale within the experimental uncertainty for the major portion of the evaporation time. Close to the end of evaporation process, we can observe the deviation from the predicted value due to precipitate formation caused due to the solute present in the liquid drop.

Evaporation of sessile drops in general occurs in distinct modes known as constant contact radius mode (CCR) and constant contact angle mode (CCA). CCR mode of evaporation represents evaporating sessile drops where contact radius is held fixed due to pinned three phase contact line. In CCR mode the drop evaporates and the reduction of drop volume is accompanied by a reduction of drop height and contact angle. In contrast CCA mode represents evaporating sessile drop for a fixed contact angle and reduction of contact radius. For sessile drops with acute contact angle and pinned contact line, evaporation primarily occurs in CCR mode for major fraction of the evaporation time. The mode of evaporation switches from CCR to CCA at the very end of evaporation process when the drop height and contact angle becomes negligible and the drop almost becomes like a thin film (Figure 2, inset III) which retracts in CCA mode. The initial contact angle of the droplet on the glass slide is nearly 20°-40°. As expected on glass, evaporation proceeds in the constant contact mode (CCR).[36] Once the droplet becomes excessively thin due to evaporation (see Figure 2, inset II), its volume is difficult to estimate. The end of the experimental data points in Figure 2 and Figure 3 in sessile configuration does not correspond to the end of evaporation, which is confirmed only using the top view (Figure 2, inset III). The lifetime of the sessile droplet is nearly 100-120 s (within experimental errors). The Bond number for a typical sessile drop in our experiments is given by Bo ~ O ($10^{-2}$), which depicts that the droplet shape can be assumed as a spherical cap for the analysis.[29]

Therefore, for a sessile drop (assuming spherical cap geometry) the rate of mass loss in CCR mode is given by [29]

$$\frac{dm_s}{dt} = -\pi D R_c c_0 (1 - RH) q(\theta)$$

(7)

where $m_s$ is the mass of the sessile drop, $R_c$ is the contact radius and $q(\theta) = 0.27\theta^2 + 1.30$, where $\theta$ is the contact angle in radians. For the contact angles in our experiment ($\theta \approx \pi/6$), $q(\theta) \approx 1.30$ and is constant. This is due to the fact that the quadratic term of contact angle in $q(\theta)$ is negligible for contact angles smaller than $\pi/3$. Therefore, the above equation reduces to

$$\frac{dm_s}{dt} = -1.30 \pi D R_c c_0 (1 - RH)$$

(8)

where $R_c$ is the contact radius. In terms of volume the above equation can be written as

$$\frac{dV_s}{dt} = -\frac{G}{\rho_f}$$

(9)

where $G = 1.30 \pi D R_c c_0 (1 - RH)$. On the Integrating the above equation with respect to time we have

$$V_s = V_{s0} - \frac{Gt}{\rho_f}$$

(10)

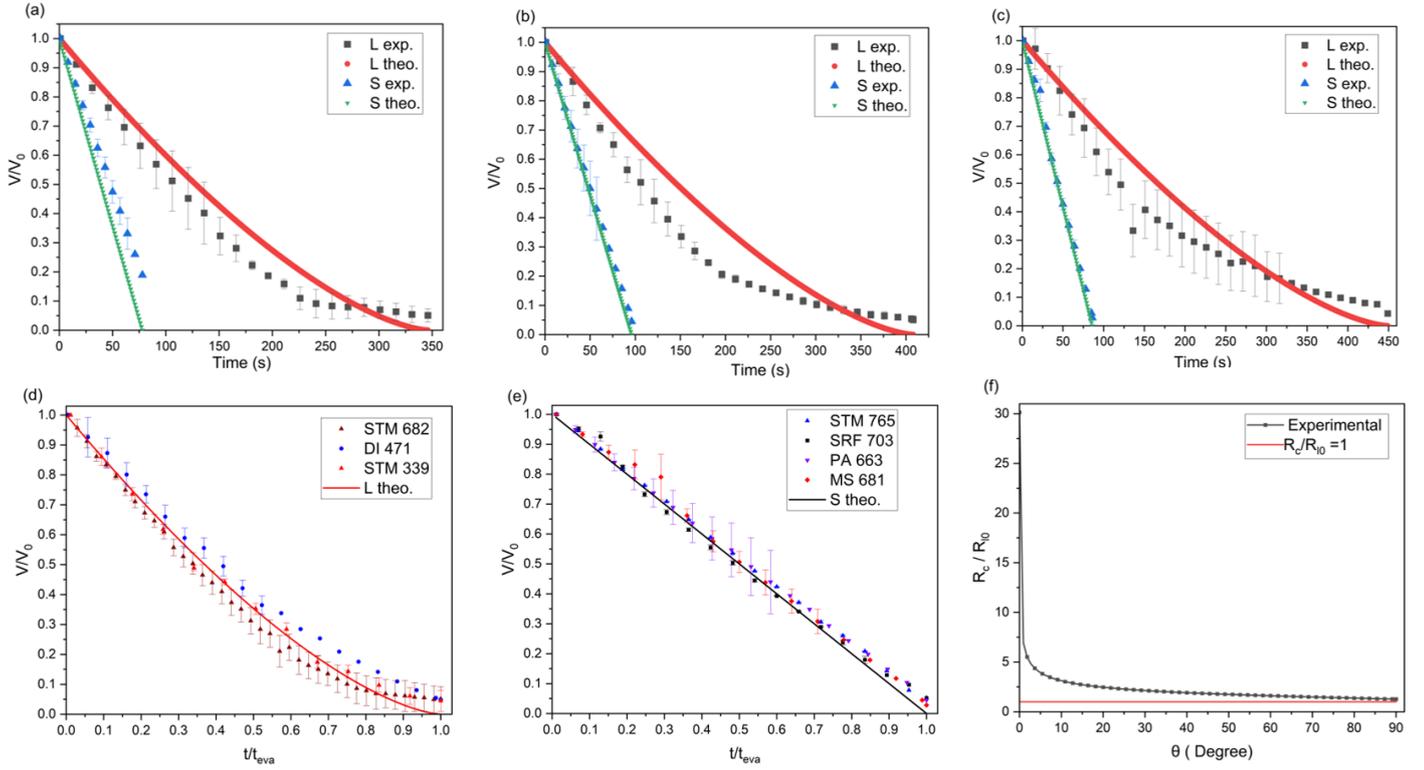

**Figure 3:** Volume ratio V/V$_0$ variation as a function of time in seconds for levitated and sessile drop evaporation for (a) KP, (b) PA, (c) MS. The theoretical model is shown is solid line and the experimental data is shown using solid markers. Here L, S denotes levitated and sessile configuration respectively. The keywords exp. and the. represents the experimental and theoretical scales of the volume regression. (d) Volume ratio V/V$_0$ versus normalize time t/t$_{eva}$ for levitated configuration: STM (initial drop diameter: 682 microns), STM (initial drop diameter: 339 microns), DI water (initial drop diameter: 471 microns). (e) Volume ratio V/V$_0$ versus normalize time t/t$_{eva}$ for sessile configuration SRF (initial contact radius:703 microns), STM (initial contact radius:765 microns), MS (initial contact radius:681 microns), PA (initial contact radius:663 microns). The theoretical model is shown is solid line and the experimental data is shown using solid markers. (f) $R_c/R_{l0}$ versus contact angle $\theta$; $R_c$ depicts the contact radius of sessile drop and $R_{l0}$ depicts the initial drop radius of leviated drop.

The instantaneous drop volume represented as a fraction of its initial volume therefore becomes

$$\frac{V_s(t)}{V_{s0}} = 1 - \frac{t}{t_{eva,s}}$$
(11)

where $t_{eva,s} = \rho_f V_0/G = \rho_f V_0/(1.30\pi D R_c c_0(1-RH))$. Comparing the instantaneous drop volume variation as a function of time we can observe, that volume regression is linear in time for sessile in comparison to nonlinear volume regression for levitated configuration. Figure 2(a) and 2(b) also shows the comparison of the drop evaporation theoretical model (green solid curve) with the experimental data in the sessile configuration for STM WT and SRF. We observe that the experimental volume regression ratio agrees with the theoretical model within the experimental uncertainty. Figure 3(a-c) shows the volume regression ratio for the bacteria laden drops (KP, PA, MS) in both sessile and levitated configuration. Figure 3(d) and 3(e) shows the normalized volume ratio $V/V_0$ as a function of normalized time coordinates $t/t_{eva}$ for drops of varying size for levitated and sessile configuration respectively. It is important to note that the evaporation process occurs similarly for drops of different sizes (both small and large).

Further the difference in evaporation time for the same initial volume for both sessile and levitated could be understood based on the mass loss rate equation (1) and (8). The instantaneous rate of mass loss for levitated configuration is proportional to the radius of the drop which itself is not constant and decreases with time due to evaporation. For sessile drop evaporation, the instantaneous rate of mass loss is proportional to the contact radius and hence remains constant. This is due to the fact that in sessile condition, 95% of the evaporation which is the major phase occur in constant contact radius mode (CCR). For levitated configuration the maximum initial rate of mass loss is always smaller than the corresponding sessile drop evaporation. This could be understood as follows:

The maximum rate of mass loss for the levitated configuration occurs at the initial condition as the drop radius is largest initially. Therefore, the initial rate of mass loss of levitated configuration has to be compared with that of sessile mode.

For sessile condition, the relationship between contact radius $R_c$ and drop volume $V_0$ assuming spherical cap is given by

$$V_0 = \frac{1}{3}\pi R_c^3 A(\theta) \tag{12}$$

where $A(\theta) = cosec^3\theta(2 - 3cos\theta + cos^3\theta)$, $\theta$ is the initial drop contact angle, $R_c$ is the contact radius. For the levitated condition assuming a sphere, the relationship between drop volume and radius is $V_0 = (4/3)\pi R_{l0}^3$. Comparing the volume ratio in both the geometries (sessile and levitated) for the same initial drop volume $V_0$ provides a scale of the ratio of contact radius in sessile mode to drop radius for levitated. The ratio is given as

$$R_c/R_{l0} = (4/A(\theta))^{1/3} > 1 \tag{13}$$

The above ratio is always greater than for acute contact angle of the drop as can be seen in Figure 3(f). For $RH = 0.42$, the ratio of evaporation rates of sessile and levitated is given as $0.55 R_c/R_{l0}$. We, therefore, observe the rate of mass loss is greater in sessile in comparison to levitated for a fixed initial volume (condition is satisfied for $R_c/R_{l0} > (1/0.55) = 1.8$) for quasi-steady diffusion limited drop evaporation with initial contact angle approximately below 40 degrees (Figure 2(d)) (ratio $R_c/R_{l0} > 1.8$ for initial contact angle below 40 degrees) as is the case for most sessile drop experiments on wetting hydrophilic surface like glass. This could also be observed from the volume ratio equation (equation 5 and 11) and drop regression visualized in $V/V_0$ and normalized time coordinates $t/t_{eva}$. We can therefore observe that drop evaporation visualized in normalized coordinates for various initial drop sizes, base fluids, and bacteria type occurs in similar way.

The difference in evaporation time in sessile and levitated configuration, in combination to the flow field results in different survivability (infectivity) of the bacteria. There are three major aspects that relates bacterial survivability with the evaporation processes.

Firstly, SRF contains various kinds of nutrients for the bacteria to consume during the evaporation. Larger time scale of evaporation raises the probability for the bacteria to absorb more nutrients. Therefore, bacteria present in levitated drop has more time to absorb nutrients than the bacteria in the sessile drop, which results in higher survivable time and hence higher infectivity.

Second important factor that relates bacterial survivability to the evaporation processes is related to evaporative flux and stresses. The geometry of sessile drop results in a non-uniform evaporative flux around the drop surface. In contrast the evaporative flux in the levitated configuration is uniform over the drop surface. The radius of the drop in the levitated configuration shrinks in size in contrast to fixed contact radius sessile drop. This results in a high constant value of evaporative mass loss in sessile in comparison to a decaying mass loss in case of levitated configuration throughout the drop evaporation phase. The decaying evaporative stress in levitated mode in comparison to constant high evaporative stresses in sessile mode results in higher survivability of bacteria in levitated configuration. Further, we also speculate that the evaporative mass loss is related to bacterial deactivation due to osmotic pressure gradient across the bacterium cell walls/membranes.

Third important factor that relates bacterial survivability to the evaporation processes is related to the fluid flow field and bacterial assembly/packing as the drop evaporates. The packing is majorly dictated by the flow field rather than diffusion processes due to very high value of Peclet number (see later section Micro-characterization of dried samples for detailed discussion on Peclet number). The non-uniform evaporative flux in sessile mode of evaporation causes a directional capillary flow towards the outer pinned edge of the drop. [37] This results in a higher fraction of the bacteria residing closer to the edge of the drop. The flow velocities generated due to capillary flow inside sessile drop is of the order of μm/s and hence we observe almost uniform bacterial deposition. In contrast the evaporative flux in the levitated configuration is uniform and the majority of the flow field generated within a levitated drop is due to acoustic streaming effects that generates toroidal flows within the drop. [38,39] The typical flow velocities inside a levitated drop are of the order of mm/s-cm/s.

The Peclet number of the bacteria is very high indicating negligible diffusion in both sessile and levitated mode. However, in the levitated phase we observe a formation of shell-like structure due to the regressing surface of the drop. As the drop evaporates and is near its complete evaporation (more than 95% volume regression has occurred), the packing efficiency/geometrical packing order is very uniform for sessile in comparison to levitated. The irregular packing along with high toroidal flow velocities makes majority of the bacteria to be at equal footing in terms of the nutrient coating in comparison to sessile condition and hence increases the chance of survivability.

## Micro-characterization of dried samples

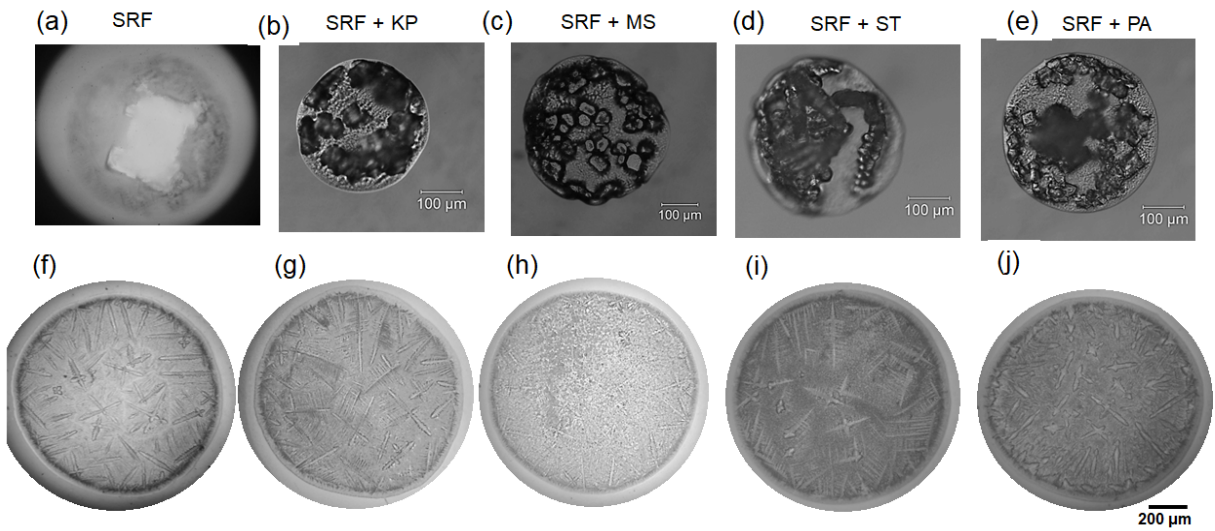

**Figure 4** Brightfield microscopy of dried sample. Levitated samples: (a) SRF (b) KP (c) MS (d) ST and (e) PA. (a-e) The scale bar represents 100 µm. (f)-(j) Sessile samples are shown in the same sequence below. (f-j) The scale bar represents 200 µm.

Levitation and sessile evaporation samples are carefully studied under a bright-field microscope (see Experimental methods). Given the three-dimensional nature of the levitated samples, the microscope was adjusted to focus on its central plane.

Figure 4 shows the brightfield microscopy images of the dried samples of SRF and bacteria laden drop in levitated (Figure 4 (a-e)) and sessile configurations (Figure 4 (f-j)). The cross-section view of the levitated sample in Figure 4a shows a crystalline (salt) center ensconced within a softer, amorphous mucin shell, indicating that salt and mucin migrate in opposite directions within the levitated droplet. The sessile samples show a thick outer edge, while the inner center is formed of dendrites observed in drying mucin-salt droplets. [40] The lack of significant distinctions between precipitates of SRF and pathogen-laden droplets is consistent with their respective drying dynamics (see Figure 2).

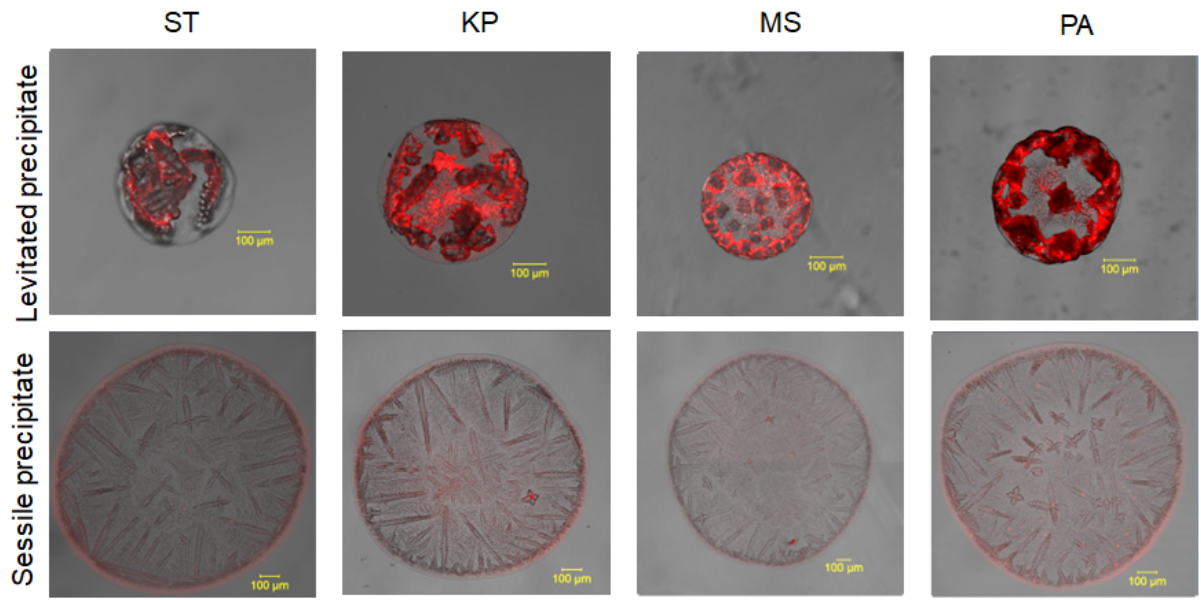

**Figure 5 Confocal microscopy of levitated (top row) and sessile (bottom row) for different pathogen laden SRF droplets.**

To gain better insights into the distribution dynamics of the pathogens in precipitates of both modes, pathogens infused with fluorescent dyes (mCherry expression for *S*. Typhimurium sample) were subjected to the same evaporation experiments as before (see Methods). Based on the fluorescent map in Figure 5 (top row), the levitated sample pathogens seem immobilized within the salt crystals. At the same time, the mucin is nearly empty of pathogens, which is consistent with the findings of Basu et al. [20] and Rasheed et al. [40]. The phenomena of bacterial distribution in an evaporating levitated droplet can be better explained by mass Peclet number analysis [20,41]:

$$Pe_m = \frac{U\, r_0}{D_{bacteria}} \tag{14}$$

Where U is the rate of droplet diameter reduction (~ 1.56 μm/s), $r_0$ = 325 μm is the initial droplet radius, and $D_{bacteria}$ is the mass diffusivity of bacteria in SRF (assuming bacterial transport similar to particle transport), which is calculated using the Stokes-Einstein equation:

$$D_{bacteria} = \frac{k_B\, T}{6\,\pi\,\mu\, r_{bacteria}} \tag{15}$$

Where $k_B$ is the Boltzmann constant, T = 302 K is the ambient temperature, μ = 1.23 × 10$^{-3}$ Pa s is the dynamic viscosity of SRF, $r_{bacteria}$ = 1 μm is the assumed bacterial radius.

We get $D_{bacteria}$ = 1.79 × 10$^{-13}$ m$^2$/s, leading to $Pe_m$ = 2.82 × 10$^3$, Hence $Pe_m \sim O(10^3)$. Here, $Pe_m \gg 1$ implies that bacteria do not diffuse in the droplet while evaporating, but it accumulates near the receding interface of the droplet. Thus, pathogens and salt form an agglomeration superstructure during evaporation in the case of levitated samples, but this work does not investigate the kinetics.

The bottom row of Figure 5 shows a near-uniform distribution of pathogens in the sessile samples. A slightly brighter outer edge indicates the pathogens' preferential migration but does not reflect the mechanics involved.

The radial velocity of the fluid flow is considered dominant in the evaporating sessile droplet, which is of the order ~ μm/s. [21] The Peclet number for the sessile droplet, Pe >> 1, suggests bacterial deposition at the droplet's contact line. [20,42] Therefore, the bacteria prominently get deposited at the edge of the evaporating sessile droplet precipitate due to capillary flow. However, since SRF also contains mucin as a constituent, the bacterial deposition is scattered over the droplet with a slightly brighter edge, which can be attributed to the polymer matrix formation of mucin. [21]

**Levitated precipitation facilitates increased survival of bacteria in comparison to the sessile precipitation**

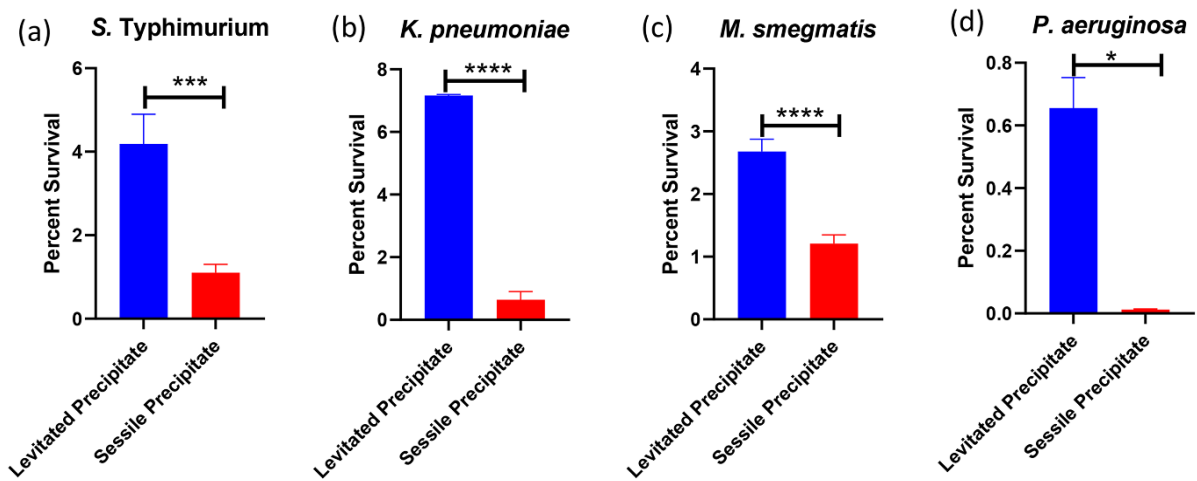

Figure 6 *In vitro* survival of four different- (a) *Salmonella* Typhimurium, (b) *Klebsiella pneumoniae*, (c) *Mycobacterium smegmatis* and (d) *Pseudomonas aeruginosa* in levitated (blue bar) or sessile precipitate (red bar) droplet. The data is representative of N=4 (biological replicates, n=4 (technical replicates).

To assess the survival of the four different bacteria (*Salmonella* Typhimurium, *Klebsiella pneumoniae*, *Pseudomonas aeruginosa*, and *Mycobacterium smegmatis*) in the sessile or levitated droplet, we performed *in vitro* survival assay by plating the reconstituted droplets onto LB agar plates (see Methods). We observed increased bacterial survival in the levitated precipitate compared to the sessile precipitate in the case of all four different bacteria (Figure 6a-d).

## Bacteria within levitated precipitate exhibit increased virulence properties in infected RAW264.7 macrophages

Given their higher survivability, we wanted to evaluate the relative infectivity in the case of either the levitated or the sessile sample. We used RAW264.7 macrophages [see Methods] to assess their respective properties. The levitated bacterial precipitate was phagocytosed less in comparison to its sessile counterpart by the RAW264.7 murine macrophages (Figure 7a). Interestingly, though the bacteria reconstituted from levitated precipitate show increased survival, the sessile droplet-laden bacteria depicted increased phagocytosis in RAW264.7 macrophages. Previous reports have suggested that phagocytes can detect several physical properties of their target, such as size, geometry, and topology, which influence the target uptake. [43] Additionally, the target rigidity determines the phagocyte response. [44] Further, the macrophages have the ability to discriminate between the live and the dead bacteria while phagocytosis. [45]

However, upon infection, the levitated bacterial precipitate showed enhanced intracellular proliferation within the infected macrophages when compared to the sessile bacterial precipitate (Figure 7b). The results were consistent for all four bacterial samples. The increased number of viable bacteria in the levitated precipitate might be causing increased infection in the RAW264.7 cells. On the other hand, the increased number of non-viable or dead bacteria in the sessile precipitate could trigger increased phagocytic uptake by the RAW264.7 cells [46] and eventually show reduced intracellular proliferation.

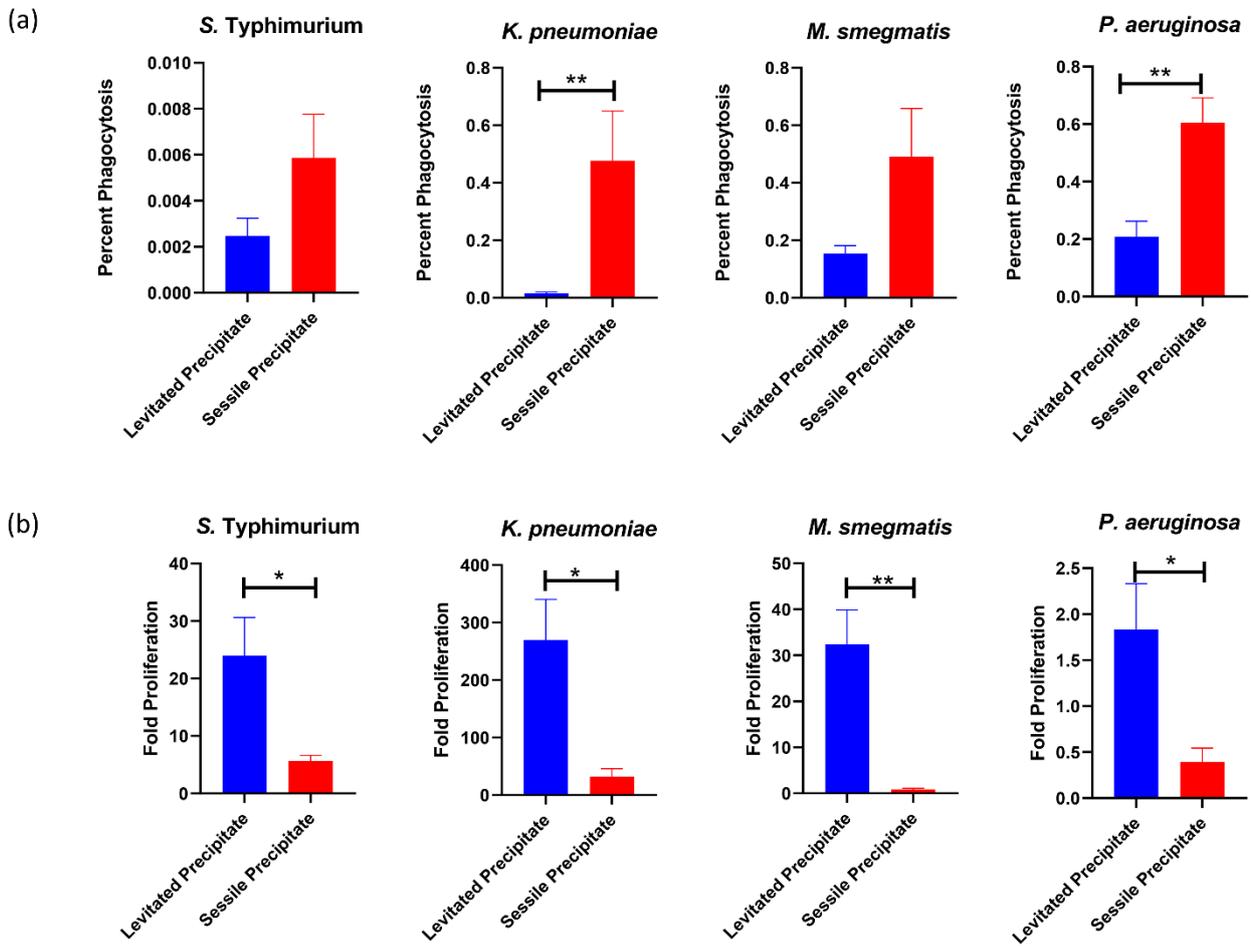

**Figure 7** Percent phagocytosis (a) and intracellular proliferation (b) of four different bacteria namely *Salmonella* Typhimurium, *Klebsiella pneumoniae*, *Mycobacterium smegmatis* and *Pseudomonas aeruginosa* retrieved from levitated (blue bar) or sessile precipitate (red bar) droplet for infection within murine RAW264.7 macrophages. The data is representative of N=3 (biological replicates, n=3 (technical replicates).

**Mode of evaporation influences the viability of the bacteria within the droplet with sessile precipitate harbouring an increased number of non-viable, and dead bacteria.**

Our previous findings suggested an increased number of dead and non-viable bacteria in the sessile precipitate compared to the levitated precipitate. To investigate our conjecture, we estimated the generation of Reactive Oxygen Species (ROS) using 2',7'Dichlorofluorescein diacetate (DCFDA) staining via flow cytometry. DCFDA is a chemically reduced form of fluorescein used as an indicator of ROS in cells. ROS cascade is one of the lethal stressors that leads to microbial cell death. [47,48] It is hypothesized that the bacteria in sessile droplets are exposed to higher evaporation-induced stress leading to higher ROS generation, which might ultimately lead to their death. Around 56.07% of *S.* Typhimurium from the sessile precipitate showed increased ROS production compared to only 30.44% of the *S.* Typhimurium population in the levitated precipitate (Figure 8a). Similarly, 83.33 % of *K. pneumoniae* in the sessile precipitate exhibited high ROS generation when compared to the levitated precipitate population of only 74.51% (Figure 8b). Only 12.07% of the *M. smegmatis* population produced increased ROS in the levitated condition. In contrast, around 42.64% of the sessile precipitate population showed heightened ROS generation (Figure 8c). However, in the case of *P. aeruginosa*, both the levitated and the sessile precipitate showed comparable ROS generation (Figure 8d). Moreover, with propidium iodide (PI) mediated live-dead staining of the bacteria in flow cytometry, we observed an increased percentage of PI-positive dead bacteria in the sessile precipitate of all four bacteria in comparison to the levitated precipitate (Figure 9 a-d). PI is a fluorescent probe used to probe dead cells in a population as it is selectively uptaken by dead cells, and live cells are impermeant to this dye.

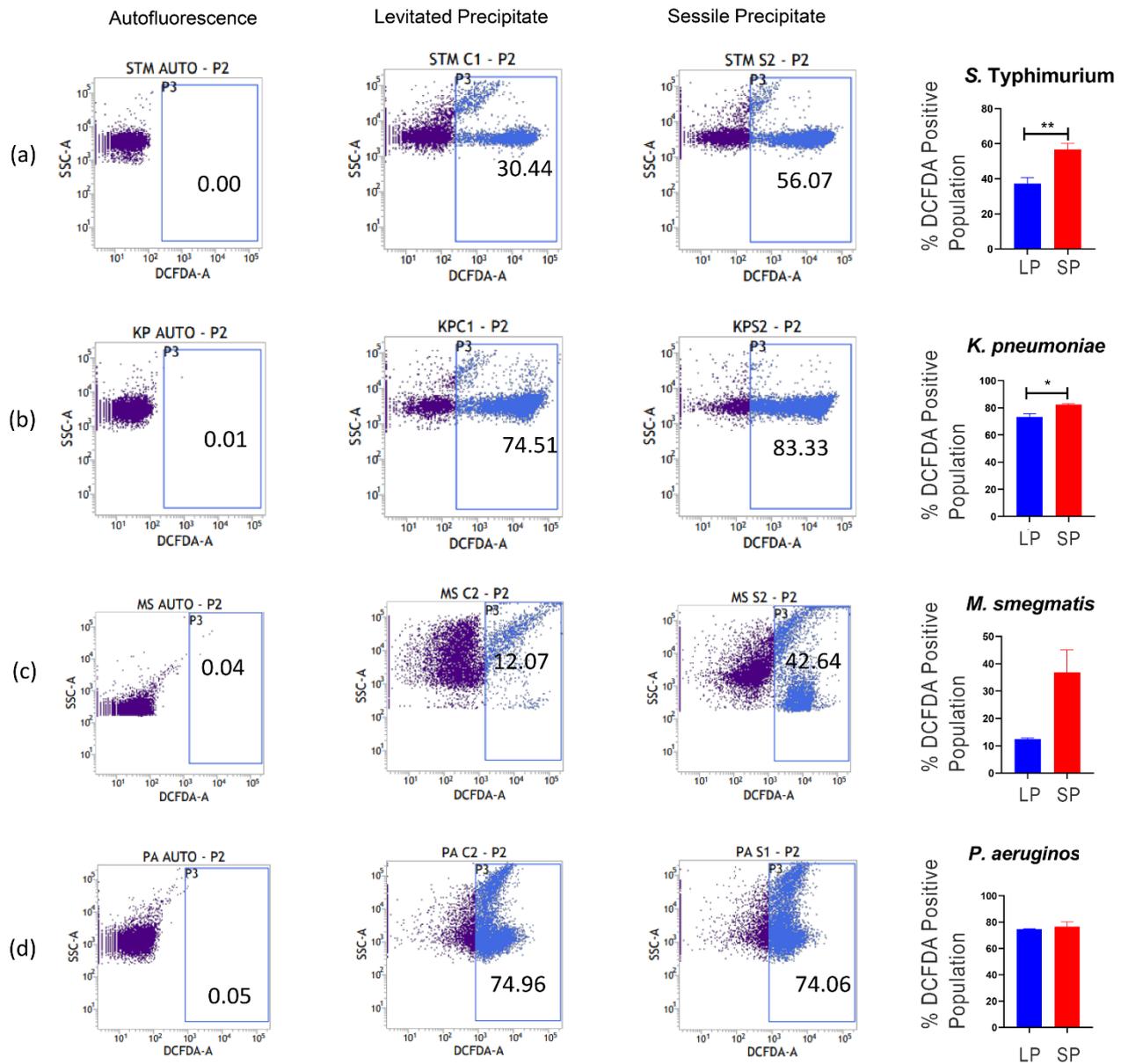

**Figure 8** Flow cytometric quantitation of Reactive oxygen species (ROS) of four different-*Salmonella* Typhimurium (a), *Klebsiella pneumoniae* (b), *Mycobacterium smegmatis* (c) and *Pseudomonas aeruginosa* in levitated (blue bar) or sessile precipitate (red bar) droplet. The data is representative of N=3 (biological replicates, n=2 (technical replicates). Here LP is Levitated precipitate and SP is sessile precipitate.

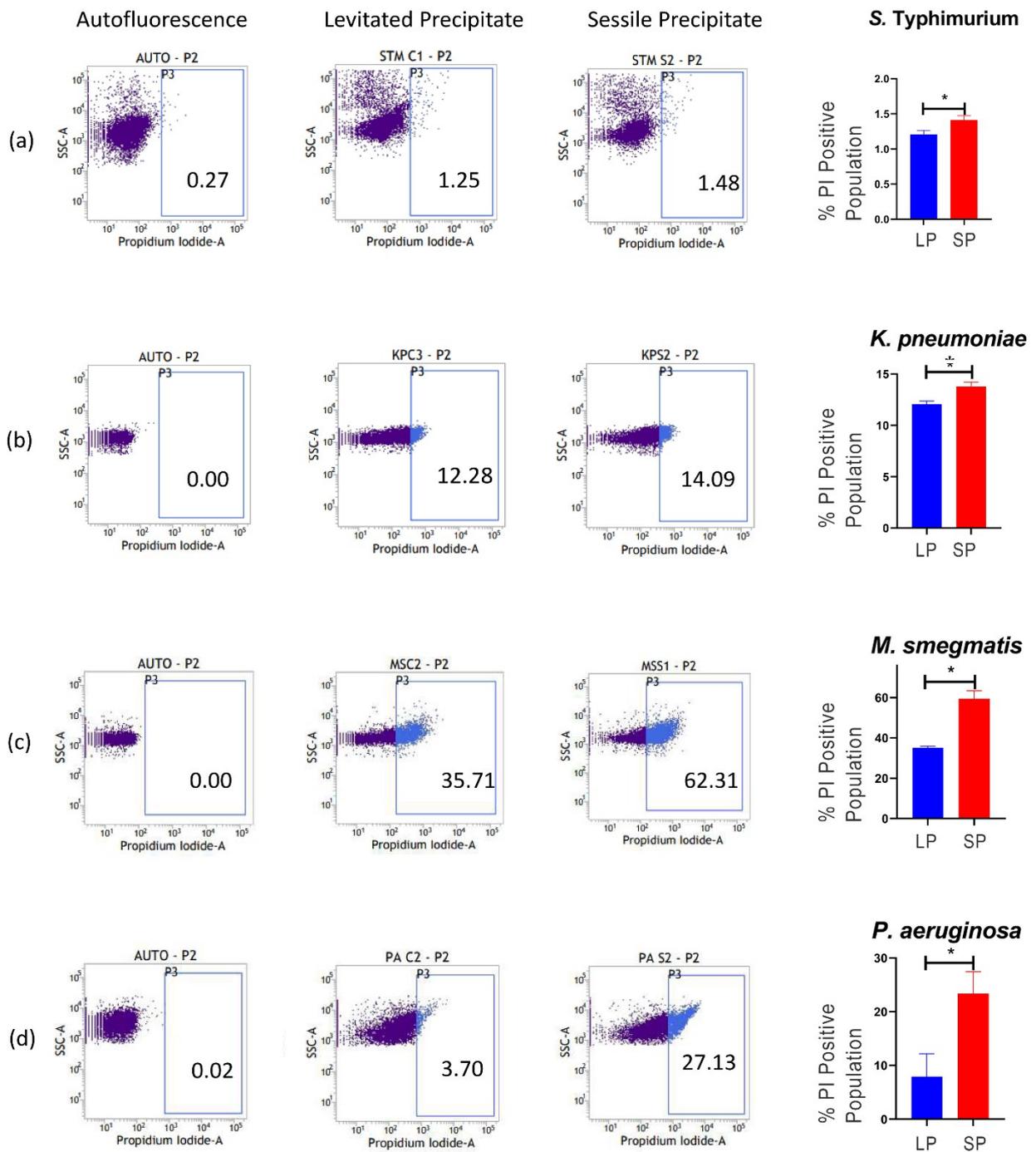

**Figure 9** Live-dead quantification by Propidium iodide (PI) staining in levitated and sessile bacterial precipitates of four different-*Salmonella* Typhimurium (a), *Klebsiella pneumoniae* (b), *Mycobacterium smegmatis* (c) and *Pseudomonas aeruginosa* in levitated (blue bar) or sessile precipitate (red bar) droplet. The data is representative of N=3 (biological replicates, n=2 (technical replicates). Here LP is Levitated precipitate and SP is sessile precipitate.

## The viable bacteria in the levitated precipitate compensate for the infection inefficiency of the sessile precipitate in the mixed infection studies

The preceding discussion established that an increased presence of stressed and dead bacteria in the sessile precipitate might be attributed to their increased phagocytic uptake by the infected RAW264.7 macrophages and decreased intracellular proliferation. In lay terms, this implies that the fomites are less likely to infect than inspired levitated droplets. A more likely scenario is that a person could encounter both aerosols and fomites; there are no insights on how this could alter the infectivity of a bacteria. We hypothesized that the live bacteria present within the levitated precipitate could overcome the infection and replication defect of the sessile-precipitated bacteria. To do so, we infected RAW264.7 macrophages with a 1:1 ratio of levitated and sessile-precipitated bacteria. We observed decreased phagocytosis of the mixed culture population similar to that of the levitated precipitated bacteria (Figure 10a). Similarly, the intracellular proliferation of the mixed infection sample was increased to that of the levitated precipitated bacterial sample (Figure 10b). These results were replicated in the confocal imaging study wherein the *S.* Typhimurium expressing mCherry (STM-RFP) levitated bacteria, and *S.* Typhimurium expressing GFP (STM-GFP) sessile precipitated bacteria were used for infecting RAW264.7 macrophages alongside 1:1 ratio of sessile- STM-GFP and levitated-STM-RFP. We assessed the bacterial phagocytic uptake and their intracellular survival via confocal microscopy at 2hr and 16hr post-infection, respectively, in STM-RFP (levitated) and STM-GFP (sessile) mono-infection and co-infection (1:1) scenario. Our result in the co-infection study revealed increased intracellular proliferation of STM-RFP (levitated) in comparison to STM-GFP (sessile) despite STM-RFP being phagocytosed less (Supplementary Figure 1-2). Altogether, our results depict that the viable bacteria in the levitated bacterial precipitate could overcome the infection defect of the sessile precipitate in the mixed infection studies.

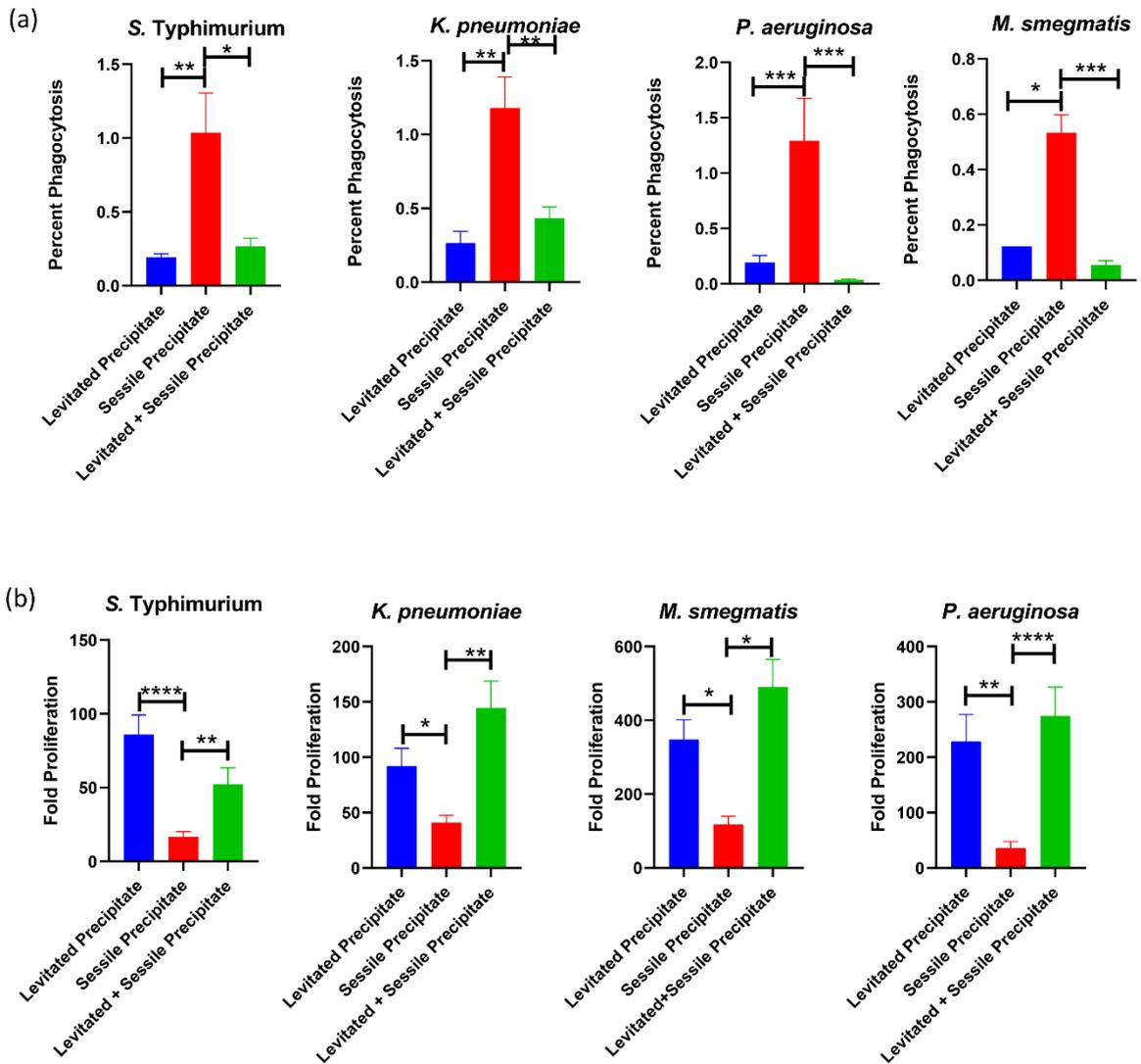

**Figure 10** Percent phagocytosis (a) and intracellular proliferation (b) of four different bacteria namely *Salmonella* Typhimurium, *Klebsiella pneumoniae*, *Mycobacterium smegmatis* and *Pseudomonas aeruginosa* within murine RAW264.7 macrophages upon retrieval from levitated (blue bar) or sessile precipitate (red bar) droplet or 1:1 ratio of levitated and sessile precipitated bacteria for infection within murine RAW264.7 macrophages. The data is representative of N=2 (biological replicates, n=2 (technical replicates).

# Conclusions

We show that desiccation dynamics play a pivotal role in bacterial survival and virulence. For the same initial volume of the droplet, the average mass evaporation rate for the sessile droplet is at least one order larger than the corresponding levitated sample demonstrating more evaporative stress experienced by the bacteria in the sessile than in levitated droplet.

The confocal images of the pathogenic levitated samples show the pathogens surrounding the salt crystals in a scattered form; since the mass Peclet number for levitated samples is $Pe_m \gg 1$, the pathogens accumulate near the receding edge of the droplet and follow inhomogeneous distribution in the precipitate. For the sessile droplet, the pathogens migrate toward the contact line of the droplet due to capillary driven flow.

The *in vitro* comparative viability studies revealed increased survival of all four different bacteria in the levitated droplet in compared to the sessile precipitate. Further, bacteria within levitated precipitate exhibited increased virulence properties in infected RAW264.7 macrophages than in the sessile precipitate with increased intracellular survival. We observed increased non-viable, dead bacteria in the sessile precipitate alongside increased ROS, which triggered increased phagocytic uptake by the RAW264.7 macrophages [46] and eventually showed reduced intracellular proliferation. Therefore, our study shows that the mode of evaporation influences the viability of the bacteria within the droplet, with sessile precipitate harbouring an increased number of non-viable or dead bacteria. The viable bacteria in the levitated precipitate possess the capability to compensate for the infection inefficiency of the sessile precipitate in the mixed infection studies. From these experimental conclusions, our study deduces that a recipient (animal/human, etc.) ingesting the precipitate of bacteria-laden droplet in a contact-free environment is more prone to get infected than the precipitate of the bacteria-laden droplet on a hydrophilic substrate/fomite as the levitated droplets emanating out of the patients while sneezing or coughing or talking may be the prime source of active infection.

## Supplementary Material

Table S1: Bacterial fluid property chart. Averaged viscosity for the shear rate range of 10-1000 1/s at 25 º C. Density and Surface tension data acquired at 29 ± 2 º C temperature and 42 ± 2% RH

Figure S1: Intracellular survival of *S.* Typhimurium present within different precipitated droplet in infected RAW264.7 macrophages at 2 hr post infection (N=2, n>50)

Figure S2: Intracellular survival of *S.* Typhimurium present within different precipitated droplet in infected RAW264.7 macrophages at 16 hr post infection (N=2, n>50)

Section S1: Brief introduction on p values and statistical analysis.

## Acknowledgements

This work was supported by the DAE SRC fellowship (DAE00195) and DBT-IISc partnership umbrella program for advanced research in biological sciences and Bioengineering to DC. Infrastructure support from ICMR (Centre for Advanced Study in Molecular Medicine), DST (FIST), and UGC (special assistance) is highly acknowledged along with ASTRA- Chair fellowship, TATA Innovation grant, and DBT-IOE partnership grant to DC. SB acknowledges financial support from the Science and Engineering Research Board (SERB). ANA gratefully acknowledges the Prime Minister's Research Fellows (PMRF) fellowship scheme facilitated by the Ministry of Higher Education (MHRD), Government of India. DH sincerely acknowledges the CSIR- SPM fellowship for her financial support. VJ appreciates the C.V.

## Author Contribution

DC, SB, DH, and ANA have conceptualized the study. DH and ANA have contributed to the experiment designing, visualization, methodology, investigation, formal analysis, literature survey, validation, writing (original draft), reviewing, and editing of the manuscript. DR has extensively done the data analysis and modelling for the evaporation dynamics. VJ has performed the experiment, methodology, visualization, and investigation along with DH and ANA. PK and DR extensively took part in writing-reviewing and editing the manuscript. DC and SB have contributed to the funding acquisition, project administration, and overall work supervision. All authors took part in writing-reviewing and editing the manuscript and approved the final version of the manuscript.

## Abbrevations

DC – Prof. Dipshikha Chakravortty

SB – Prof. Saptarshi Basu

DH – Dipasree Hajra

ANA- Amey Nitin Agharkar

VJ – Dr. Vivek Jaiswal

PK – Dr. Prasenjit Kabi

DR- Dr. Durbar Roy

# Conflict of Interest

The authors have declared that no conflict of interest exists.

# Data Availability Statement

The data that supports the findings of this study are available within the article and its supplementary material.

**Address correspondence to**: Prof. Saptarshi Basu, 406, Department of Mechanical Engineering, Indian Institute of Science (IISc), Bangalore 560012, India, Telephone number: +91 80 22933367, Email: sbasu@iisc.ac.in